\documentclass[10pt]{iopart}

\usepackage{iopams,bm,cite,color,graphicx,mathptmx,url}

\newcommand{\openone}{\leavevmode\hbox{\small1\normalsize\kern-.33em1}}
\newcommand{\binom}[2]{{#1 \choose #2}}
\eqnobysec

\begin{document}

\title{Quantum polarization characterization and tomography}

\author{J~S\"{o}derholm$^{1,2}$, G~Bj\"{o}rk$^{3,2}$,
  A~B~Klimov$^{4,2}$,  L~L~S\'{a}nchez-Soto$^{1,5,2}$ and G~Leuchs$^{1,6}$}

\address{$^{1}$ Max-Planck-Institut f\"ur die Physik des Lichts,
  G\"{u}nther-Scharowsky-Stra{\ss}e 1, Bau 24, 91058 Erlangen,
  Germany}

\address{$^2$ NORDITA, Roslagstullsbacken 23,
SE-106 91 Stockholm, Sweden}

\address{$^3$ Department of Applied Physics,
 Royal Institute of Technology (KTH), AlbaNova University Center,
 SE-106 91 Stockholm, Sweden}

\address{$^4$ Departamento de F\'{\i}sica, Universidad
 de Guadalajara, 44420 Guadalajara, Jalisco, Mexico}

\address{$^5$ Departamento de \'{O}ptica, Facultad de F\'{\i}sica,
 Universidad Complutense, 28040 Madrid, Spain}

\address{$^{6}$ Institut f\"ur Optik, Information und Photonik,
  Staudtstra{\ss}e 7, 91058 Erlangen, Germany}

\ead{gbjork@kth.se}

\begin{abstract}
 We present a complete polarization characterization of any
 quantum state of two orthogonal polarization modes,  and give a
 systematic measurement procedure to collect the necessary data.
 Full characterization requires measurements of the photon number
 in both modes and linear optics. In the situation where only the
 photon-number difference can be determined, a limited but useful
 characterization is obtained. The characteristic Stokes moment
 profiles are given for several common quantum states.
\end{abstract}

\pacs{03.65.Wj, 42.50.Dv, 42.25.Ja, 03.65.Ta}



\section{Introduction}

Far from its source, any freely propagating electromagnetic field can
be considered to a good approximation as a plane wave, with its
electric field lying in a plane perpendicular to the direction of
propagation. This simple observation is the root of the notion of
polarization. At first  glance, it may seem rather obvious how to
translate such a concept into the realm of quantum optics. However,
hurdles such as hidden polarization~\cite{Klyshko}, the fact that the
Poincar\'{e} sphere is too small to accommodate states with excitation
larger than one photon~\cite{Muller}, and the difficulties in defining
polarization properties of two-photon entangled
fields~\cite{Sergienko}, to cite only a few examples, show that the
classical theory, mainly based on first-order polarization moments,
is insufficient for quantized fields.

Here, we outline a systematic method for polarization characterization
of quantum fields. The method is based on a simple premise; namely,
that if we can predict the $m$th-order moment of the Stokes operator
in any direction on the Poincar\'{e} sphere, we know all there is to
be known  about the state polarization of this order~\cite{BjorkPRA},
including any correlations between the Stokes operators.
A tensor representation of the polarization information is based on
such correlations.
However, expressing the Stokes moments as functions of the measurement
directions gives a more compact representation and provides a natural
visualization. The Stokes profile representation also gives a
relevant characterization for passive interferometry. Our analysis
below makes use of both representations.

As a state's polarization properties do not require the full density
matrix to be determined, it allows polarization tomography to be
more easily performed than full quantum tomography
\cite{RaymerQCM,RaymerPRA,Karassiov}. Considering polarization tomography
with ideal detection, we show that the number of measurement directions
can be made equal to the number of independent parameters.

The remaining material of the article is organized as follows.
After recalling the fundamentals on the quantum description of
polarization in section~\ref{Sec:Background}, we present our scheme
for characterization of quantum polarization properties in
section~\ref{Sec:HigherOrder}. In section~\ref{Sec:Tomography}, we
consider how the necessary data can be obtained experimentally.
We thus arrive at an efficient method, which is feasible for
polarization tomography of few-photon states. In
section~\ref{Sec:Menagerie}, we apply our characterization to
several classes of states. Finally, our conclusions are
presented in section~\ref{Sec:Conclusions}.

\section{Setting the scene}
\label{Sec:Background}

In the following, we consider monochromatic plane waves. Such fields
can be decomposed into two orthogonal
transverse modes, such as the horizontally and vertically polarized
modes. For highly focused beams or waves in a waveguide, the
plane-wave description is often inadequate, as the field is not longer
transverse. It is our belief that the concepts discussed in this paper
can also be extended to
such non-plane waves, and several proposals have already appeared in
the literature \cite{Carozzi,Setala,Luis3D}. However, we shall not
discuss such generalizations here.

The classical theory for the polarization of plane waves was
established by Stokes already more than 150 years ago~\cite{Stokes}.
We shall build on his theory as the basis of our treatment will be
the Stokes operators, whose expectation values are the Stokes
parameters~\cite{Collett}.  Following the conventions used in the
quantum theory of angular momentum~\cite{Schwinger} and in quantum
optics~\cite{Yurke,LuisPO}, we define the Stokes operators as
\begin{equation}
  \begin{array}{ll}
    \hat{S}_{0} = \hat{a}_{H}^\dagger \hat{a}_{H}
    + \hat{a}_{V}^\dagger \hat{a}_{V} \, ,  \qquad  &
    \hat{S}_{1} = \hat{a}_{H} \hat{a}_{V}^\dagger
    + \hat{a}_{H}^\dagger \hat{a}_{V} \, , \\
    & \\
    \hat{S}_{2} = i ( \hat{a}_{H} \hat{a}_{V}^\dagger -
    \hat{a}_{H}^\dagger \hat{a}_{V} ) \, ,   &
    \hat{S}_{3} = \hat{a}_{H}^\dagger \hat{a}_{H} -
    \hat{a}_{V}^\dagger \hat{a}_{V} \, ,
  \end{array} \label{Stokop}
\end{equation}
where $\hat{a}_{H}$ and $\hat{a}_{V}$ are the annihilation operators
of the modes associated with horizontally and vertically oscillating
fields, respectively. With this choice, the usual ordering of the
Stokes parameters $\mathcal{I} = \langle \hat{S}_{0} \rangle$,
$\mathcal{Q} = \langle \hat{S}_{3} \rangle$,
$\mathcal{U} = \langle \hat{S}_{1} \rangle$, and
$\mathcal{V} = \langle \hat{S}_{2} \rangle$ differs from that of the
indices of the operators.  However, as far as the theory below is
concerned, we could just as well have associated any other pair of
orthogonal polarization modes to these operators. That would only
influence the interpretation of the theory and not the theory itself.

As the annihilation and creation operators obey the bosonic
commutation relations $[ \hat{a}_{\alpha}, \hat{a}_{\beta}^\dagger ] =
\delta_{\alpha \beta}$, for $\alpha, \beta \in \{ H, V \}$, the Stokes
operators satisfy the commutation relations of an su(2) algebra
\begin{equation}
  \label{ccrsu2}
  [ \hat{S}_{j} , \hat{S}_{k} ] = i 2 \, \varepsilon_{j k \ell} \hat{S}_{\ell} \, ,
\end{equation}
where the latin indices run from 1 to 3 and $\varepsilon_{j k \ell}$
is the fully antisymmetric Levi-Civita tensor. The noncommutability of these
operators precludes the simultaneous exact measurement of the corresponding
physical quantities. The variances
$(\Delta S_j)^2 = \langle \hat{S}_j^2 \rangle - \langle \hat{S}_j \rangle^2$
are found to obey the uncertainty relation
\begin{equation}
  2 \langle \hat{S}_{0} \rangle \leq
  (\Delta S_{1})^2 + (\Delta S_{2})^2 + (\Delta S_{3})^2
  \leq  \langle  \hat{S}_{0} (\hat{S}_{0} + 2) \rangle  \, .
  \label{UncertRelation}
\end{equation}
Moreover, while the Stokes operators are all Hermitian, the noncommutability
makes ``mixed,'' non-symmetric products (such as $\hat{S}_{1} \hat{S}_{2}$)
non-Hermitian, also precluding their direct measurement.

The standard definition of the degree of polarization for a quantum
state $\hat{\varrho}$ is
\begin{equation}
  \label{P1}
  \mathbb{P}_S (\hat{\varrho})  =
  \frac{| \langle \hat{\mathbf{S}} \rangle |}
  {\langle \hat{S}_0 \rangle} =
  \frac{\sqrt{\langle \hat{S}_1 \rangle^2
      + \langle \hat{S}_2 \rangle^2
      +   \langle \hat{S}_3 \rangle^2}}
  {\langle \hat{S}_0 \rangle} \, ,
\end{equation}
where $\hat{\mathbf{S}} = (\hat{S}_1, \hat{S}_2, \hat{S}_3)$ and
$\langle \hat{\mathbf{S}} \rangle$ is the Stokes vector.
Note that only first-order moments of the Stokes operators
are used in this definition. In a more elaborated characterization,
the degree of polarization can be subdivided into excitation manifolds
according to the total photon number $N$.  This makes physical sense
because since the corresponding observable $\hat{S}_0$ commutes with
all the other Stokes operators
\begin{equation}
  [\hat{S}_0 , \hat{S}_j] = 0 \, ,
\end{equation}
a complete set of simultaneous eigenstates of $\hat{S}_0$ and any of
$\hat{S}_1$, $\hat{S}_2$, and $\hat{S}_3$ does exist.  In fact, the
statistics of the latter three operators is usually determined by a
set of wave plates, a polarizing beam splitter, and two
photodetectors, giving (in the ideal case) information not only about
$\hat{S}_1$, $\hat{S}_2$, or $\hat{S}_3$, but simultaneously of
$\hat{S}_0$.

Let us here take a quick look at excitation manifolds $N = 1$ and $N =
2$.  One can readily convince oneself that any pure single-photon
state $| \Psi_1 \rangle$ satisfies $\mathbb{P}_S (| \Psi_1 \rangle) =
1$, i.e., any such state is fully polarized according to the
definition (\ref{P1}).  In fact, for an arbitrary single-photon state
$\hat{\varrho}_{1}$, the degree of polarization is related to the
purity $\Tr ( \hat{\varrho}^2 ) $ according to
\begin{equation}
  \mathbb{P}_S (\hat{\varrho}_1) = \sqrt{2 \,
    \Tr ( \hat{\varrho}_{1}^{2} ) - 1} \, .
  \label{P1purity}
\end{equation}
However, this relation does not hold for other excitation manifolds.
For example, any pure state in excitation manifold $N = 2$ of the
form~\cite{BjorkOC}
\begin{equation}
  | \Psi (a, \theta) \rangle =
  a e^{-i \theta} |2,0 \rangle + i \sqrt{1 - 2 a^2} \, |1,1 \rangle
  + a e^{i \theta} |0,2 \rangle \, ,
  \label{2phUnpol}
\end{equation}
where $a$ and $\theta$ are real numbers and $0 \leq a \leq
1/\sqrt{2}$, satisfies $\mathbb{P}_S (| \Psi (a, \theta) \rangle) =
0$, which indicates that it is unpolarized.  However, as we shall see
below, these states have polarization structure (they are not
isotropic in the polarization sense) and cannot be regarded as
unpolarized.

\section{Higher-order polarization properties}
\label{Sec:HigherOrder}

In order to characterize the polarization properties of a state, we shall
employ measurements of higher-order moments of the Stokes operators.
This is very close to the Glauber correlation functions in quantum
coherence theory~\cite{Glauber}, and has common grounds with Klyshko
generalized coherence matrices~\cite{Klyshko}.  In a recent paper
\cite{BjorkPRA}, we have used the central moments for higher-order
polarization characterization.  Whereas the central moments may be
preferred by some readers, the raw moments used in the present work
seem to allow for an easier and more systematic approach.

As we have already discussed, one can perform a measurement of the
total photon number without disturbing the measurement of any other
Stokes operator.  In classical optics, this is tantamount to the fact
that the state of polarization is independent of the intensity.  This
suggests that the polarization properties are given by
$\hat{\mathbf{S}}$.  However, an ideal measurement of polarization
provides some information about the total energy and vice versa.  For
example, an even (odd) measured eigenvalue of any of the observables
$\hat{S}_1$, $\hat{S}_2$, and $\hat{S}_3$ implies an even (odd) total
number of photons.  Also, determining the probability distribution for
the total number of photons $p_N$ simultaneously sets bounds on the
polarization properties in accordance with the inequalities
(\ref{UncertRelation}).

Taking these observations into account, we distinguish polarization
properties for different numbers of photons, and let full polarization
characterization refer to complete knowledge of the expectation values
of all possible combinations of the Stokes operators.  The $r$th-order
polarization information of a state $\hat{\varrho}$ is then given by
$p_N$ and the expectation values of the form
\begin{equation}
  T_{j_1 j_2 \ldots j_r}^{(r,N)} =
  \langle \hat{S}_{j_1} \hat{S}_{j_2} \ldots \hat{S}_{j_r} \rangle_N =
  \Tr ( \hat{\varrho}_N \hat{S}_{j_1} \hat{S}_{j_2} \ldots  \hat{S}_{j_r} ) \, ,
  \label{rthOrderExpectationValues}
\end{equation}
where $j_k \in \{ 1 , 2, 3 \}$ and $\hat{\varrho}_{N}$ denotes the
normalized two-mode, $N$-photon state obtained by projecting
$\hat{\varrho}$ onto the $N$th excitation manifold
\begin{equation}
  \hat{\varrho}_N = \frac{\hat{\openone}_{N} \hat{\varrho}
  \hat{\openone}_{N}}{p_{N}} \, . \label{rhoN}
\end{equation}
Using the Fock basis, the projector can thus be expressed as
$\hat{\openone}_N = \sum_{n=0}^N | n , N - n \rangle \langle n , N - n |$
and $p_{N} = \Tr (\hat{\openone}_{N} \hat{\varrho})$.  For any given order
$r$ and excitation manifold $N$, the elements
(\ref{rthOrderExpectationValues}) form a Cartesian tensor
$\mathbf{T}^{(r,N)} (\hat{\varrho})$ of rank $r$.  Due to the
Hermiticity of the Stokes operators, theses tensors satisfy
\begin{equation}
  T_{j_1 \ldots j_r}^{(r,N)}  = [T_{j_r \ldots j_1}^{(r,N)} ]^\ast \, .
  \label{TensorHermiticity}
\end{equation}
We leave $\hat{\varrho}_N$ and $\mathbf{T}^{(r,N)}$ undefined for
any $N$ such that $p_N = 0$, and employ the convention that they then
do not contribute to sums.

When $(\mu,\nu,j_k)$ is a cyclic permutation of $(1,2,3)$,
the commutation relation~(\ref{ccrsu2}) implies that polarization
tensor elements of neighboring ranks are related according to
\begin{equation}
  T_{j_{1} \ldots \mu \nu \ldots j_{r-1}}^{(r,N)} -
  T_{j_{1} \ldots \nu  \mu \ldots j_{r-1}}^{(r,N)}  =
  i 2 \, T_{j_{1} \ldots j_{k} \ldots j_{r-1}}^{(r-1,N)} \,  .
  \label{TensorElementRelation}
\end{equation}
Hence, $\mathbf{T}^{(r-1,N)}$ can be determined from
$\mathbf{T}^{(r,N)}$ and, consequently, $\mathbf{T}^{(R,N)}$
determines all $\mathbf{T}^{(r,N)}$ such that $r < R$. Complete
polarization information of order $R$ is thus equivalent to complete
polarization information of all orders $r \leq R$.

Using the relations (\ref{Stokop}) and (\ref{ccrsu2}), it is also
straightforward to show that the polarization information carried by
$\mathbf{T}^{(R,N)}$ is equivalent to that contained in the set of
generalized coherence matrices of orders $2 r$ ($r \leq R$), whose
elements are of the form~\cite{Klyshko} $ \langle
(\hat{a}_{H}^\dagger)^j (\hat{a}_{V}^\dagger)^{r-j} \hat{a}_{H}^k
\hat{a}_{V}^{r-k} \rangle_N$.  Having complete polarization
information of all orders about a state $\hat{\varrho}$ is therefore
equivalent to knowing its block-diagonal
projection~\cite{BjorkOC,RaymerQCM,Karassiov,Korolkova}
\begin{equation}
  \hat{\varrho}_{\mathrm{pol}} =
  \sum_{N=0}^\infty p_{N} \hat{\varrho}_{N} \, ,
  \label{Brho} \\
\end{equation}
where $\hat{\varrho}_N$ is given by (\ref{rhoN}). This is the so
called the polarization sector (or polarization density matrix).
The number of parameters characterizing a block-diagonal state
limited to the excitation manifolds $N_1,N_2,\ldots,N_\nu$ is
\begin{equation}
 - 1 + \sum_{k=1}^\nu (N_k + 1)^2 . \label{ParametersForArbitraryManifolds}
\end{equation}
In particular, when a state is limited to the manifolds
$0,1,\ldots,\widetilde{N}$, the number of parameters simplifies to
\begin{equation}
 - 1 + \sum_{N=0}^{\widetilde{N}} (N + 1)^2 = \frac{\widetilde{N}
 (2 \widetilde{N}^2 + 9 \widetilde{N} + 13)}{6} . \label{ParametersBrho}
\end{equation}
For such a state, complete polarization information of order
$\widetilde{N}$ is sufficient to determine its block-diagonal
projection (\ref{Brho}).  The general density matrix for a state with
no more than $\widetilde{N}$ photons is determined by $\widetilde{N}
(\widetilde{N} + 3) (\widetilde{N}^2 + 3 \widetilde{N} + 4)/4$
independent real numbers. Hence, the polarization share of this
information quickly decreases with $\widetilde{N}$.

\section{Polarization tomography}
\label{Sec:Tomography}

We now turn to the question of how to characterize polarization
properties experimentally.  Since our interest is limited to the
information contained in the block-diagonal projection~(\ref{Brho}),
it is clear that we are not required to do full quantum state
tomography~\cite{RaymerQCM,Karassiov,RaymerPRA}.  As complete
polarization information corresponds to doing quantum tomography of
all $N$-photon Hilbert spaces excited by the considered state, one can
make use of the methods developed for finite-dimensional
systems~\cite{Newton,Leonhardt,Weigert}.  However, some recently
proposed higher-order intensity measurements~\cite{Schilling} seem to
be closest related to the ones we present below.

We will assume ideal measurements and that the total photon number and
its probability distribution $p_N$ can be determined.  This is
obviously a severe restriction apart from the lowest excitation
manifolds.  However, the situation where no information about the
total photon number can be obtained is described by simply summing
over the different manifolds as discussed in
section~\ref{Sec:PolPropAlone}.

Below, we also treat the experimental determination of different
moments of an observable as different measurements. In principle,
each moment requires an infinite number of measurement runs in order
to be determined exactly.  This would also give us the full
probability distribution of the eigenvalues and thus all the moments.
However, for the vast majority of realistic probability distributions,
a lower moment requires fewer runs to be accurately determined.

\subsection{Moment measurements}
\label{Sec:MomentMeasurements}

The fact that the classical Stokes parameters are easily determined
experimentally makes them highly practical.  Also in quantum optics,
the measurement setups corresponding to the fundamental Stokes
operators are simple.  These setups are composed only by phase
shifters, beam splitters and photon-number measurements.  The effects
of linear optical devices are described by SU(2)
transformations~\cite{Yurke}, which can be expressed as
\begin{equation}
 \hat{U} (\Phi,\Theta,\Xi) = e^{-i \Phi \hat{S}_3/2}
 e^{-i \Theta \hat{S}_2/2} e^{-i \Xi \hat{S}_3/2} ,
 \label{GeneralSU2Transformation}
\end{equation}
where $\Phi$, $\Theta$, and $\Xi$ are the Euler angles.
Any such transformation can be easily realized using linear optics~\cite{Simon}
and they are lossless, so they leave $\hat{S}_0$ unaffected.

Let us now introduce the Stokes operator in an arbitrary direction
characterized by the unit vector $\mathbf{n}  \in \mathbb{R}^{3}$ as
\begin{equation}
 \hat{S}_\mathbf{n} \equiv \mathbf{n} \cdot \hat{\mathbf{S}} =
 \sum_{k=1}^3 n_k \hat{S}_k .
\end{equation}
The effect of an arbitrary SU(2) transformation on $\hat{S}_\mathbf{n}$,
can then be expressed as
\begin{eqnarray}
 \hat{U} (\Phi,\Theta,\Xi) \, \hat{S}_\mathbf{n} \, \hat{U}^\dagger
 (\Phi,\Theta,\Xi) = \hat{S}_{\mathbf{R}_3 (\Phi) \cdot
 \mathbf{R}_2 (\Theta) \cdot \mathbf{R}_3 (\Xi) \cdot \mathbf{n}} ,
 \qquad \label{USnUdagger}
\end{eqnarray}
where $\mathbf{R}_k (\phi)$ denotes the matrix describing a rotation of
$\phi$ around the $\mathbf{e}_k$-axis, e.g.
\begin{equation}
 \mathbf{R}_1 (\phi) = \left[ \begin{array}{ccc} 1 & 0 & 0 \cr 0 &
 \cos \phi & - \sin \phi \cr 0 & \sin \phi & \cos \phi \end{array} \right] .
\end{equation}
Hence, any SU(2) transformation corresponds to a proper rotation in
$\mathbb{R}^3$ \cite{Yurke}. We note that $\hat{S}_3$, which gives
the photon-number difference, is transformed according to
\begin{equation}
 \fl
 \hat{U} (\Phi,\Theta,\Xi) \, \hat{S}_3 \, \hat{U}^\dagger (\Phi,\Theta,\Xi)
 = \hat{S}_{\mathbf{n}} , \quad \quad \quad
 \mathbf{n} = (\sin \Theta \cos \Phi , \sin \Theta \sin \Phi , \cos \Theta) .
 \label{US3Udagger}
\end{equation}
That is, $\hat{U} (\Phi,\Theta, 0)$ is the standard displacement on the sphere
and the transformation parameters $\Theta$ and $\Phi$ equal the spherical
coordinates of the vector $\mathbf{n}$ characterizing the transformed
Stokes operator. We see that any $\hat{S}_\mathbf{n}$ is related to
$\hat{S}_3$ by an SU(2) transformation corresponding to a polarization
rotation of $\Theta/2$ followed by a differential phase shift of $\Phi/2$.
Expectation values of the form $\langle \hat{S}_\mathbf{n}^r \rangle$ can
thus be straightforwardly determined experimentally.
Ideally, we can simultaneously measure the total photon number $N$,
so that also expectation values of the form
$\langle \hat{S}_\mathbf{n}^r \rangle_N$ can be determined.

The tensor $\mathbf{T}^{(r,N)}$ gives any expectation value of the
form
\begin{equation}
  \langle \hat{S}_{\mathbf{n}_1} \hat{S}_{\mathbf{n}_2}  \ldots
  \hat{S}_{\mathbf{n}_r} \rangle_N   =
  \sum_{j_1=1}^3 \ldots \sum_{j_r=1}^3
  n_{j_1}^{(1)} \ldots n_{j_r}^{(r)} T_{j_1 \ldots j_r}^{(r,N)} \, .
  \label{GeneralSumOverT}
\end{equation}
When all vectors are the same, (\ref{GeneralSumOverT}) simplifies
considerably. For a given state, the relation between
$\langle \hat{S}_\mathbf{n}^r \rangle_N$  and the direction
$\mathbf{n}$ will be referred to as the $N$-photon Stokes moment
profile of order $r$. These profiles can be expressed as
\begin{equation}
   \langle \hat{S}_\mathbf{n}^r \rangle_N = \sum_{k=0}^r
   \sum_{\ell=0}^{r-k}
   n_1^k n_2^\ell n_3^{r-k-\ell} M_{k,\ell}^{(r,N)} \ , \label{SnrMean}
\end{equation}
where the moment component $M_{k,\ell}^{(r,N)}$ is the sum of all
tensor elements of the form $T_{j_1 \ldots j_r}^{(r,N)}$ that have
$k$ ones and $\ell$ twos as subscripts. Due to
(\ref{rthOrderExpectationValues}), every moment component is thus
the expectation value of the Hermitian operator formed by the sum
of the Stokes-operator products corresponding to its tensor elements.
The number of such elements is given by the trinomial coefficient
\begin{equation}
   (k,\ell,r - k - \ell)! = \frac{r!}{k! \ell! (r - k - \ell)!} .
   \label{trinomial}
\end{equation}
For example, we have $M_ {1,1}^{(3,N)} = T_{123}^{(3,N)} +
T_{132}^{(3,N)} + T_{213}^{(3,N)} + T_{231}^{(3,N)} + T_{312}^{(3,N)}
+ T_{321}^{(3,N)}$.  We note that the sum of all tensor elements of
order $r$ in excitation manifold $N$ can be written as
\begin{equation}
   \sum_{j_1=1}^3 \ldots \sum_{j_r=1}^3 T_{j_1 \ldots j_r}^{(r,N)} =
   \sum_{k=0}^r \sum_{\ell=0}^{r-k} M_{k,\ell}^{(r,N)} =
   3^{r/2} \langle \hat{S}_{\mathbf{n}_\mathrm{diag}}^r \rangle_N  \,   ,
   \label{SumOfTensorElements}
\end{equation}
where $\mathbf{n}_\mathrm{diag} = (1,1,1)/\sqrt{3}$.

Since the polarization tensor satisfies the Hermiticity condition
(\ref{TensorHermiticity}), the moment components are real, and
consequently the real part of $\mathbf{T}^{(r,N)}$ is sufficient to
determine $\langle \hat{S}_\mathbf{n}^r \rangle_N$ in any direction
$\mathbf{n}$. Naturally, knowing the Stokes moment profile
(\ref{SnrMean}) is equivalent to knowing the
\begin{equation}
  m_r = \frac{(r + 1) (r + 2)}{2}
\end{equation}
moment components $M_{k,\ell}^{(r,N)}$.

Moreover, using the commutation relation (\ref{ccrsu2}), it is
possible to determine the differences between the tensor elements
belonging to the same moment component $M_{k,\ell}^{(r,N)}$ from
$\mathbf{T}^{(r-1,N)}$.  Since every element of $\mathbf{T}^{(r,N)}$
belongs to such a moment component, it thus follows that
$\mathbf{T}^{(r-1,N)}$ together with all $M_{k,\ell}^{(r,N)}$
determine $\mathbf{T}^{(r,N)}$.

Let us now introduce a standard ordering of the Stokes operators
according to
\begin{equation}
   \hat{\mathcal{O}}_{k,\ell}^{(r)} =
   \hat{S}_1^k \hat{S}_2^\ell \hat{S}_3^{r-k-\ell} .
\end{equation}
Making repeated use of the commutation relation (\ref{ccrsu2}), the
moment components can then be expressed as
\begin{equation}
   M_{k,\ell}^{(r,N)} = (k,\ell,r - k - \ell)! \,
   \langle \hat{\mathcal{O}}_{k,\ell}^{(r)} \rangle_N  +
   \langle \hat{\mathcal{C}}_{k,\ell}^{(r)} \rangle_N \, , \label{MOC}
\end{equation}
where $\hat{\mathcal{C}}_{k,\ell}^{(r)}$ is a sum over
Stokes-operator products of orders smaller than $r$. The
Casimir operator
\begin{equation}
 \hat{\mathbf{S}}^2 = \sum_{k=1}^3 \hat{S}_k^2 =
 \hat{S}_0 (\hat{S}_0 + 2) \label{Casimir}
\end{equation}
implies that, for $r \geq 2$, we have
\begin{equation}
 \fl
 \langle \hat{\mathcal{O}}_{k+2,\ell}^{(r)} \rangle_N  +
 \langle \hat{\mathcal{O}}_{k,\ell+2}^{(r)} \rangle_N +
 \langle \hat{\mathcal{O}}_{k,\ell}^{(r)} \rangle_N =
 N (N + 2) \langle \hat{\mathcal{O}}_{k,\ell}^{(r-2)} \rangle_N  +
 \langle \hat{S}_1^k [ \hat{S}_1^2 , \hat{S}_2^\ell ]
 \hat{S}_3^{r-k-\ell-2} \rangle_N \, ,
 \label{InvariantOrelation}
\end{equation}
where the last term can again be written as a sum over
Stokes-operator products of orders smaller than $r$.  Equations
(\ref{MOC}) and (\ref{InvariantOrelation}) show that there is a
relation between moment components of orders $r$ and $r - 2$, and
that the number of independent moment components of order $r$ is $
m_r - m_{r-2} = 2 r + 1$.  For a general $N$-photon state, the number
of independent moment components to determine is thus $\sum_{r=1}^N
(2 r + 1) = N (N + 2)$. For a general block-diagonal state, we also
have to determine the probability distribution for the total number
of photons.  Assuming that the excited manifolds are known to be
limited to $N_1,N_2,\ldots,N_\nu$, we find the number of independent
parameters to be $\nu - 1 + \sum_{k=1}^\nu N_k (N_k + 2)$, which is
in agreement with (\ref{ParametersForArbitraryManifolds}).

\subsubsection{General single-photon state}

In the basis $(|1,0 \rangle , |0,1 \rangle )$, the density matrix of a
general single-photon state can be written as
\begin{equation}
  \hat{\varrho}_1 = \left (
    \begin{array}{cc}
      \pi_0 & R + i I \cr R - i I & 1 - \pi_0
    \end{array}
  \right ) ,
\end{equation}
where $R^2 + I^2 \leq \pi_0 (1 - \pi_0)$. Using a superscript to
identify state-specific average values, the first-order Stokes moment
profile is given by
\begin{equation}
  \langle \hat{S}_\mathbf{n} \rangle_1^{\hat{\varrho}_1} =
  2 R n_1 - 2 I n_2 + (2 \pi_0 - 1) n_3 .
\end{equation}
Hence, the three moment components are seen to be independent.

\subsubsection{General two-photon state}

Using the basis $(|2,0 \rangle , |1,1 \rangle , |0,2 \rangle )$, the
density matrix of a general two-photon state can be written as
\begin{equation}
  \hat{\varrho}_2 = \left (
    \begin{array}{ccc}
      \pi_1 & R_1 + i I_1 & R_2 + i I_2 \cr R_1 - i I_1 & \pi_2 & R_3 + i I_3 \\
      R_2 - i I_2 & R_3 - i I_3 & 1 - \pi_1 - \pi_2
    \end{array}
  \right ) \, .
\end{equation}
The first- and second-order Stokes moment profiles can then be
expressed as
\begin{eqnarray}
  \langle \hat{S}_\mathbf{n} \rangle_2^{\hat{\varrho}_2} & = & 2
  \sqrt{2} \, [(R_1 + R_3) n_1 - (I_1 + I_3) n_2] + 2 \, (2 \pi_1 +
  \pi_2 - 1) n_3 , \\
  & & \nonumber \\
  \langle \hat{S}_\mathbf{n}^2 \rangle_2^{\hat{\varrho}_2} & = &
  2 \, (1 + \pi_2 + 2 R_2) n_1^2 + 2 \, (1 + \pi_2 - 2 R_2) n_2^2 +
  4 \, (1 - \pi_2) n_3^2 - 8 I_2 n_1 n_2 \nonumber \\
  & + & 4 \sqrt{2} \, n_3 [(R_1 - R_3) n_1 - (I_1 - I_3) n_2] ,
  \label{2ph2ndOrder}
\end{eqnarray}
which makes it easy to identify the moment components.
As implied by (\ref{Casimir}), the moment components of the three
first terms of (\ref{2ph2ndOrder}) are determined by two parameters.
Hence, there are only five independent second-order moment components.

\subsection{Choosing measurement directions}

We have seen that the information content of the moment components
allows us to do polarization tomography by only measuring moments.
Performing the moment measurements in increasing order, the $2 r + 1$
independent moment components for each order $r$ and manifold $N$ can
be determined by choosing equally many directions $\mathbf{n}$ such
that (\ref{SnrMean}) gives linearly independent equations for the
unknown moment components.

\subsubsection{First order}

Quite naturally, both the first-order moment components and the
first-order polarization tensors are given by the manifold-specific
Stokes parameters
\begin{equation}
  M_{\delta_{1 j},\delta_{2 j}}^{(1,N)} = T_j^{(1,N)} =
  \langle  \hat{S}_j \rangle_N \, . \label{FirstOrderMTS}
\end{equation}
The three sets of information are hence identical and are obtained by
determining the expectation value $\langle \hat{S}_\mathbf{n}
\rangle_N$ for the directions $\mathbf{n} = (1,0,0)$, $(0,1,0)$, and
$(0,0,1)$.  As the operators $\hat{S}_\mathbf{n}$ and
$\hat{S}_{-\mathbf{n}}$ only differ by the signs of their eigenvalues,
the corresponding measurements will give the same information.  Hence,
equivalent measurements correspond to a line through the origin.
Choosing three orthogonal directions as above thus results in a
uniform distribution of the measurements on the Poincar\'{e} sphere.

\subsubsection{Second order}
\label{Sec:SecondOrder}

We have seen that there are five independent moment components of
second order. Thinking of the measurements as lines, we choose the
directions as
\begin{equation}
  \fl  \displaystyle
  \mathbf{n}_{1,2} = \frac{(0 , \pm 2 , 1 + \sqrt{5})}
  {\sqrt{10 + 2 \sqrt{5}}} ,
  \qquad
  \mathbf{n}_{3,4} = \frac{(\pm 2 , 1 + \sqrt{5} , 0)}
  {\sqrt{10 + 2 \sqrt{5}}} ,
  \qquad
  \mathbf{n}_5 = \frac{(1 + \sqrt{5} , 0 , 2)}
  {\sqrt{10 + 2 \sqrt{5}}} \, ,
\end{equation}
which maximizes the minimum angle between the
lines~\cite{FejesToth,Conway} and thus in some sense spreads out the
measurements over the Poincar\'{e} sphere as much as possible.  The
six second-order moment components are then given by
\begin{eqnarray}
  \fl
  M_{0,1}^{(2,N)} = \frac{\sqrt{5}}{2}
  \langle \hat{S}_{\mathbf{n}_1}^2 - \hat{S}_{\mathbf{n}_2}^2 \rangle_N , \\
  \fl
  M_{1,1}^{(2,N)} = \frac{\sqrt{5}}{2} \langle \hat{S}_{\mathbf{n}_3}^2
  -  \hat{S}_{\mathbf{n}_4}^2 \rangle_N , \\
  \fl
  M_{1,0}^{(2,N)} = \frac{\sqrt{5}}{2}  \langle \hat{S}_{\mathbf{n}_1}^2
  +  \hat{S}_{\mathbf{n}_2}^2 + \hat{S}_{\mathbf{n}_3}^2 +
  \hat{S}_{\mathbf{n}_4}^2  + 2 \hat{S}_{\mathbf{n}_5}^2 \rangle_N
  - \sqrt{5} N (N + 2) , \label{M102N} \\
  \fl
  M_{0,0}^{(2,N)} = \frac{(15 + 7 \sqrt{5})
  \langle  \hat{S}_{\mathbf{n}_1}^2 +
  \hat{S}_{\mathbf{n}_2}^2 \rangle_N
    - (10 + 4 \sqrt{5}) \langle \hat{S}_{\mathbf{n}_3}^2 +
   \hat{S}_{\mathbf{n}_4}^2 \rangle_N + (6 + 2 \sqrt{5}) N (N + 2)}{4 (7 + 3 \sqrt{5})} , \\
  \fl
  M_{0,2}^{(2,N)} = \frac{(10 + 4 \sqrt{5}) \langle
   \hat{S}_{\mathbf{n}_1}^2
  +  \hat{S}_{\mathbf{n}_2}^2 \rangle_N + (25 + 11 \sqrt{5}) \langle
  \hat{S}_{\mathbf{n}_3}^2 +
  \hat{S}_{\mathbf{n}_4}^2 \rangle_N - (14 + 6 \sqrt{5}) N (N + 2)}{4 (7 + 3 \sqrt{5})} , \\
  \fl
  M_{2,0}^{(2,N)} = \frac{(36 + 16 \sqrt{5}) N (N + 2) -
   (25 + 11 \sqrt{5}) \langle \hat{S}_{\mathbf{n}_1}^2 +
   \hat{S}_{\mathbf{n}_2}^2 \rangle_N - (15 + 7 \sqrt{5}) \langle
   \hat{S}_{\mathbf{n}_3}^2 +
   \hat{S}_{\mathbf{n}_4}^2 \rangle_N}{4 (7 + 3 \sqrt{5})} .
\end{eqnarray}
Note that their determination does not require any first-order
measurement.  The second-order polarization tensors can be expressed
in the first- and second-order moment components as
\begin{eqnarray}
  \mathbf{T}^{(2,N)} =
  \left (
    \begin{array}{ccc}
      M_{2,0}^{(2,N)} &
      \frac{M_{1,1}^{(2,N)}}{2} + i \langle \hat{S}_3 \rangle_N &
      \frac{M_{1,0}^{(2,N)}}{2} - i \langle \hat{S}_2 \rangle_N \\
      \frac{M_{1,1}^{(2,N)}}{2} - i \langle \hat{S}_3 \rangle_N &
      M_{0,2}^{(2,N)} &
      \frac{M_{0,1}^{(2,N)}}{2} + i \langle \hat{S}_1 \rangle_N \\
      \frac{M_{1,0}^{(2,N)}}{2} + i \langle \hat{S}_2 \rangle_N &
      \frac{M_{0,1}^{(2,N)}}{2} + i \langle \hat{S}_1 \rangle_N &
      M_{0,0}^{(2,N)}
    \end{array}
  \right ) .
  \label{2ndOrderTensorM}
\end{eqnarray}
When writing tensors, we let larger entities and rows correspond to
tensor indices placed to the left of those corresponding to smaller
entities and columns.

As an aside, we decompose the Stokes-operator covariance matrix into
different excitation manifolds
$\bm{\Gamma} = \sum_{N=0}^\infty p_N \bm{\Gamma}_N$
and note that the matrix elements are given by
\begin{eqnarray}
 \fl
 \Gamma_{jk,N} \equiv \frac{\langle \hat{S}_j \hat{S}_k \rangle_N +
 \langle \hat{S}_k \hat{S}_j \rangle_N}{2} -
 \langle \hat{S}_j \rangle_N \langle \hat{S}_k \rangle_N =
 \mathrm{Re} ( T_{jk}^{(2,N)} ) - T_j^{(1,N)} T_k^{(1,N)} .
\end{eqnarray}
For any state that satisfies $\mathbb{P}_S = 0$, we thus have
$\bm{\Gamma}_N = \mathrm{Re} ( \mathbf{T}^{(2,N)} )$.

\subsubsection{Third order}
\label{Sec:ThirdOrder}

We know that there are seven independent
third-order moment components.  Maximizing the minimum angle between
seven lines~\cite{Conway}, we find that the measurements should
correspond to $\hat{S}_1$, $\hat{S}_2$, $\hat{S}_3$, and the
directions
\begin{equation}
  \mathbf{n}_{4,5} = \frac{(\pm 1 , 1 , 1)}{\sqrt{3}} ,
  \quad \quad \quad
  \mathbf{n}_{6,7} = \frac{(\pm 1 , -1 , 1)}{\sqrt{3}} .
\end{equation}
However, this choice gives only four independent measurements, since
we have
\begin{equation}
  \hat{S}_1^3 = \frac{3 \sqrt{3}}{8} \, ( - \hat{S}_{\mathbf{n}_4}^3 +
  \hat{S}_{\mathbf{n}_5}^3 -  \hat{S}_{\mathbf{n}_6}^3 +
  \hat{S}_{\mathbf{n}_7}^3 )
  + \frac{3 N (N + 2) - 4}{2} \, \hat{S}_1
\end{equation}
and similar relations for $\hat{S}_2^3$ and $\hat{S}_3^3$.  By
choosing directions close to $\hat{S}_1$, $\hat{S}_2$, and
$\hat{S}_3$, it is possible to determine all third-order moment
components.  However, this choice would make it hard to obtain the
necessary data, since the corresponding expectation values differ only
slightly from the known $\langle \hat{S}_1^3 \rangle_N$, $\langle
\hat{S}_2^3 \rangle_N$, and $\langle \hat{S}_3^3 \rangle_N$.
Consequently, although highly symmetric polyhedrons have been
successfully applied in protocols for tomography of multi-qubit
states \cite{Bogdanov}, the related method considered here fails.
How to optimally choose the measurement directions
for higher-order polarization tomography thus appears to be a
complicated problem. This notwithstanding, the third-order
polarization tensors can be expressed as
\begin{equation}
 \fl \mathbf{T}^{(3,N)} = \left ( \begingroup \everymath{\scriptstyle}
 \begin{array}{ccc}
  M_{3,0}^{(3,N)} & \frac{M_{2,1}^{(3,N)} + i 4 T_{1,3}^{(2,N)} +
  i 2 T_{3,1}^{(2,N)}}{3} & \frac{M_{2,0}^{(3,N)} - i 4 T_{1,2}^{(2,N)}
  - i 2 T_{2,1}^{(2,N)}}{3} \cr \frac{M_{2,1}^{(3,N)} -
  i 2 T_{1,3}^{(2,N)} + i 2 T_{3,1}^{(2,N)}}{3} & \frac{M_{1,2}^{(3,N)}
  + i 2 T_{2,3}^{(2,N)} + i 4 T_{3,2}^{(2,N)}}{3} &
  \frac{M_{1,1}^{(3,N)}}{6} + i T_{1,1}^{(2,N)} - i T_{2,2}^{(2,N)} +
  i T_{3,3}^{(2,N)} \cr \frac{M_{2,0}^{(3,N)} + i 2 T_{1,2}^{(2,N)} -
  i 2 T_{2,1}^{(2,N)}}{3} & \frac{M_{1,1}^{(3,N)}}{6} -
  i T_{1,1}^{(2,N)} - i T_{2,2}^{(2,N)} + i T_{3,3}^{(2,N)} &
  \frac{M_{1,0}^{(3,N)} - i 2 T_{3,2}^{(2,N)} - i 4 T_{2,3}^{(2,N)}}{3}
  \cr \hline \frac{M_{2,1}^{(3,N)} - i 2 T_{1,3}^{(2,N)} -
  i 4 T_{3,1}^{(2,N)}}{3} & \frac{M_{1,2}^{(3,N)} + i 2 T_{2,3}^{(2,N)}
  - i 2 T_{3,2}^{(2,N)}}{3} & \frac{M_{1,1}^{(3,N)}}{6} +
  i T_{1,1}^{(2,N)} - i T_{2,2}^{(2,N)} - i T_{3,3}^{(2,N)}
  \cr \frac{M_{1,2}^{(3,N)} - i 4 T_{2,3}^{(2,N)} -
  i 2 T_{3,2}^{(2,N)}}{3} & M_{0,3}^{(3,N)} & \frac{M_{0,2}^{(3,N)} +
  i 4 T_{2,1}^{(2,N)} + i 2 T_{1,2}^{(2,N)}}{3} \cr
  \frac{M_{1,1}^{(3,N)}}{6} + i T_{1,1}^{(2,N)} + i T_{2,2}^{(2,N)} -
  i T_{3,3}^{(2,N)} & \frac{M_{0,2}^{(3,N)} - i 2 T_{2,1}^{(2,N)} +
  i 2 T_{1,2}^{(2,N)}}{3} & \frac{M_{0,1}^{(3,N)} +
  i 2 T_{3,1}^{(2,N)} + i 4 T_{1,3}^{(2,N)}}{3} \cr \hline
  \frac{M_{2,0}^{(3,N)} + i 2 T_{1,2}^{(2,N)} +
  i 4 T_{2,1}^{(2,N)}}{3} & \frac{M_{1,1}^{(3,N)}}{6} -
  i T_{1,1}^{(2,N)} + i T_{2,2}^{(2,N)} + i T_{3,3}^{(2,N)} &
  \frac{M_{1,0}^{(3,N)} - i 2 T_{3,2}^{(2,N)} +
  i 2 T_{2,3}^{(2,N)}}{3} \cr \frac{M_{1,1}^{(3,N)}}{6} -
  i T_{1,1}^{(2,N)} + i T_{2,2}^{(2,N)} - i T_{3,3}^{(2,N)} &
  \frac{M_{0,2}^{(3,N)} - i 2 T_{2,1}^{(2,N)} -
  i 4 T_{1,2}^{(2,N)}}{3} & \frac{M_{0,1}^{(3,N)} +
  i 2 T_{3,1}^{(2,N)} - i 2 T_{1,3}^{(2,N)}}{3} \cr
  \frac{M_{1,0}^{(3,N)} + i 4 T_{3,2}^{(2,N)} + i 2 T_{2,3}^{(2,N)}}{3}
  & \frac{M_{0,1}^{(3,N)} - i 4 T_{3,1}^{(2,N)} -
  i 2 T_{1,3}^{(2,N)}}{3} & M_{0,0}^{(3,N)}
  \end{array} \endgroup \right) . \label{3rdOrderTensorM}
\end{equation}

\subsection{Recurrence relation for Stokes moment profiles}

Above, we have seen that the $N$ lowest-order Stokes moment profiles
$\{ \langle \hat{S}_\mathbf{n}^r \rangle_N \}_{r=1}^N$ contain all
polarization information of an $N$-photon state.  In particular, we
show in the appendix that the higher-order profiles are determined by
the recurrence relation
\begin{equation}
  \fl
  \langle \hat{S}_\mathbf{n}^{N + 1 + \mu} \rangle_N =
  \left\{
    \begin{array}{ll}
      \displaystyle - \sum_{j=1}^{N/2} 4^{N/2 + 1 - j} f (N + 2, 2 j)
      \,  \langle \hat{S}_\mathbf{n}^{2 j - 1 + \mu} \rangle_N ,
      & N \ \mathrm{even,} \\
      \displaystyle - \sum_{j=0}^{\frac{N-1}{2}} 4^{\frac{N + 1}{2} - j}
      f (N + 2, 2 j + 1) \, \langle \hat{S}_\mathbf{n}^{2 j + \mu}
      \rangle_N , & N \ \mathrm{odd,}
    \end{array}
  \right. \label{Srecurrence}
\end{equation}
where $\mu$ is a non-negative integer and $f (n,k)$ are the central
factorial numbers of the first kind given by~\cite{Butzer}
\begin{equation}
 \fl f (n,k) = \left\{
 \begin{array}{ll}
   0 , & n < k , \\
   \delta_{n,0} , & k = 0 , \\
   \displaystyle \binom{2 n - k}{k} k \sum_{j=0}^{n-k}
   \frac{(-1)^j}{j! (n+j)} \binom{2 n - 2 k}{n-k-j}
   \sum_{m=0}^j (-1)^m \binom{j}{m} \left( \frac{j}{2} - m \right)^{n-k+j}
   , & 1 \leq k \leq n .
 \end{array}
 \right .
\end{equation}
For the lowest excitation manifolds, we thus get
\begin{eqnarray}
  \begin{array}{ll}
    \fl \langle \hat{S}_{\mathbf{n}}^r \rangle_0 = 0 , &
    \langle \hat{S}_{\mathbf{n}}^r \rangle_1 = \left\{
      \begin{array}{ll}
        1 , & r \ \mathrm{even,} \\
        \langle \hat{S}_\mathbf{n} \rangle_1 , & r \ \mathrm{odd,}
      \end{array}
    \right. \\[0.6cm]
    \fl \langle \hat{S}_{\mathbf{n}}^r \rangle_2 = \left\{
      \begin{array}{ll}
        2^{r-2} \langle \hat{S}_\mathbf{n}^2 \rangle_2 , & r \ \mathrm{even,} \\
        2^{r-1} \langle \hat{S}_\mathbf{n} \rangle_2 , & r \ \mathrm{odd,}
      \end{array}
    \right. \qquad &
    \langle \hat{S}_{\mathbf{n}}^r \rangle_3 = \left\{
      \begin{array}{ll}
        \frac{9 - 3^{r} + (3^r - 1) \langle \hat{S}_\mathbf{n}^2
          \rangle_3}{8} , & r \ \mathrm{even,} \\
        \frac{(9 - 3^{r-1}) \langle \hat{S}_\mathbf{n} \rangle_3 +
        (3^{r-1} - 1) \langle \hat{S}_\mathbf{n}^3 \rangle_3}{8} ,
        & r \ \mathrm{odd.}
      \end{array}
    \right. \end{array}
\label{StokesProfiles0to3}
\end{eqnarray}
The above relations for $N = 1$ and $N = 2$ can also be established
from the general property
\begin{equation}
 \hat{S}_{-\mathbf{n}}^r = \left\{
 \begin{array}{rl}
   \hat{S}_\mathbf{n}^r , & r \ \mathrm{even,} \\
   - \hat{S}_\mathbf{n}^r , & r \ \mathrm{odd.}
 \end{array}
 \right. \label{SnrEvenOdd}
\end{equation}

\subsection{Non-resolved photon numbers}
\label{Sec:PolPropAlone}

As pointed out above, apart from the few lowest excitation manifolds, it
is difficult to distinguish different excitation manifolds
experimentally.  In case there is no information about the total
photon number available, the measured expectation values are weighted
averages over the manifolds of the form $\langle \hat{A} \rangle =
\sum_{N=0}^\infty p_N \langle \hat{A} \rangle_N$.  Due to linearity,
it is clear that the corresponding polarization tensors, Stokes moment
profiles, and moment components, which are given by
\begin{equation}
  \fl
  \mathbf{T}^{(r)} = \sum_{N=0}^\infty p_N \mathbf{T}^{(r,N)} ,
  \qquad
  \langle \hat{S}_\mathbf{n}^r \rangle =
  \sum_{N=0}^\infty p_N \langle \hat{S}_\mathbf{n}^r \rangle_N ,
  \qquad
  M_{k,\ell}^{(r)} = \sum_{N=0}^\infty p_N M_{k,\ell}^{(r,N)} ,
  \label{SumN}
\end{equation}
respectively, enjoy most of the properties of their manifold-specific
counterparts.  However, due to the factor $N (N + 2)$ appearing in
(\ref{InvariantOrelation}), the relations between moment components of
different orders depend on the excitation manifold.  In general, this
makes all photon-number-averaged moment components $M_{k,\ell}^{(r)}$
independent and $R$th-order polarization tomography then requires
$\sum_{r=1}^R m_r = R (R^2 + 6 R + 11)/6$ parameters to be determined.
However, the knowledge of the average photon number
$\langle \hat{S}_0 \rangle$ and its variance
$\langle \hat{S}_0^2 \rangle - \langle \hat{S}_0 \rangle^2$ is
sufficient to remove the redundancy for the second order, since we
then know the right-hand side of the relation
$M_{2,0}^{(2)} + M_{0,2}^{(2)} + M_{0,0}^{(2)} = \langle \hat{S}_0
(\hat{S}_0 + 2) \rangle$ obtained from (\ref{Casimir}).
With this partial knowledge about the photon distribution,
we can thus determine the second-order moment components using the
five measurement settings given in section~\ref{Sec:SecondOrder}.

Now, assume that we know that a state is limited to the first three
manifolds, i.e., that the number of photons cannot exceed two.  In
this case, the determination of $\langle \hat{S}_0 \rangle$, $\langle
\hat{S}_0^2 \rangle$, and the three lowest-order Stokes moment
profiles is sufficient for complete photon-resolved polarization
characterization.  Explicitly, we have $p_1 = 2 \langle \hat{S}_0
\rangle - \langle \hat{S}_0^2 \rangle$ and $p_2 = (\langle \hat{S}_0^2
\rangle - \langle \hat{S}_0 \rangle)/2$, which together with
(\ref{StokesProfiles0to3}) give
\begin{equation}
  \fl \displaystyle
  \langle \hat{S}_\mathbf{n} \rangle_1 =
  \frac{4 \langle \hat{S}_\mathbf{n} \rangle -
  \langle \hat{S}_\mathbf{n}^3 \rangle}
  {6 \langle \hat{S}_0 \rangle - 3 \langle \hat{S}_0^2  \rangle} \, ,
  \quad
  \langle \hat{S}_\mathbf{n} \rangle_2 =
  \frac{\langle \hat{S}_\mathbf{n}^3 \rangle -
  \langle \hat{S}_\mathbf{n} \rangle}{3} \, ,
  \quad
  \langle \hat{S}_\mathbf{n}^2 \rangle_2   =
  \frac{2 (\langle \hat{S}_\mathbf{n}^2 \rangle +
  \langle \hat{S}_0^2 \rangle -
  2 \langle \hat{S}_0 \rangle)}
  {\langle \hat{S}_0^2 \rangle - \langle \hat{S}_0 \rangle} .
\end{equation}

\section{A menagerie of states and their polarization properties}
\label{Sec:Menagerie}

We next apply the characterization developed above to some classes of
states.  In most cases, we give only the Stokes moment profiles for
the states, as these provide the most compact presentation of the
polarization properties.  It should be straightforward to obtain the
moment components and polarization tensors if needed.

Using (\ref{US3Udagger}), we can write the Stokes moment profiles of
an arbitrary state $\hat{\varrho}$ as
\begin{equation}
 \langle \hat{S}_\mathbf{n}^r \rangle_N = \mathrm{Tr} [ \hat{U}
 (\Phi,\Theta,\Xi) \, \hat{S}_3^r \, \hat{U}^\dagger
 (\Phi,\Theta,\Xi) \hat{\varrho} ] ,
\end{equation}
where the SU(2) transformation $\hat{U} (\Phi,\Theta,\Xi)$ is given
by (\ref{GeneralSU2Transformation}). Now, consider the state
$\hat{\varrho}^\prime$ obtained by applying an SU(2) transformation
to $\hat{\varrho}$ according to
\begin{equation}
 \hat{\varrho}^\prime = \hat{U} (\varphi,\vartheta,\xi)
 \hat{\varrho} \, \hat{U}^\dagger (\varphi,\vartheta,\xi) .
\end{equation}
As the trace of a product is invariant under cyclic permutations,
(\ref{USnUdagger}) ensures that the Stokes moment profiles of the
state $\hat{\varrho}^\prime$ are related to those of
$\hat{\varrho}$ by rotations. Indeed, we find that
$\langle \hat{S}_\mathbf{n}^r \rangle_N^{\hat{\varrho}^\prime}$
is obtained from
$\langle \hat{S}_\mathbf{n}^r \rangle_N^{\hat{\varrho}}$ by
rotating the latter $\xi$ around the $\mathbf{e}_3$-axis,
followed by a rotation of $\vartheta$ around the
$\mathbf{e}_2$-axis and another of $\varphi$ around the
$\mathbf{e}_3$-axis. That is, we have
\begin{equation}
 \langle \hat{S}_\mathbf{n}^r \rangle_N^{\hat{\varrho}^\prime} =
 \langle \hat{S}_{\mathbf{R}_3 (-\xi) \cdot \mathbf{R}_2 (-\vartheta)
 \cdot \mathbf{R}_3 (-\varphi) \cdot \mathbf{n}}^r
 \rangle_N^{\hat{\varrho}} . \label{SU2relationForProfiles}
\end{equation}
Naturally, the sequence of rotations appearing in
(\ref{SU2relationForProfiles}) is the inverse of the one described
above.  Because of these simple rotations, the determination of the
polarization properties of a state $\hat{\varrho}$, implicitly gives
the polarization properties of all states related to $\hat{\varrho}$
by an SU(2) transformation, although these states may appear very
different.  We note that common, passive, two-mode interferometers are
described by SU(2) transformations too.  The considerations below are
therefore relevant to interferometry.

\subsection{SU(2) coherent states}

The SU(2) coherent states are the eigenstates of the operators
$\hat{S}_\mathbf{n}$.  They are also the only states that minimize the
variance sum, i.e., that saturate the left inequality in the
uncertainty relation (\ref{UncertRelation}).  Using the spherical
coordinates (\ref{US3Udagger}), we have the eigenequation
$\hat{S}_\mathbf{n} | N ; \Theta , \Phi \rangle = N | N ; \Theta ,
\Phi \rangle$.  The $N$-photon, SU(2) coherent states are of the form
\begin{eqnarray}
  | N ; \Theta , \Phi \rangle & = & \sum_{n=0}^N e^{-i n \Phi}
  \sqrt{\binom{N}{n}} \sin^{N-n} \left ( \frac{\Theta}{2} \right )
  \cos^n \left ( \frac{\Theta}{2} \right ) \, | n , N - n \rangle
  \nonumber \\
  & = & e^{-i N \Phi/2} \, \hat{U} (\Phi,\Theta,0) \, | N , 0 \rangle .
\end{eqnarray}
Since an overall phase factor does not have any physical significance,
they are thus all related to the state $|N,0 \rangle$ by an SU(2)
transformation (\ref{GeneralSU2Transformation}). As discussed above,
such transformations correspond to simple rotations of the
Poincar\'{e} sphere, so we limit our explicit treatment to the states
$|N,0 \rangle$.  Making use of well-known results for the beam
splitter, we easily find the Stokes moment profiles to be
\begin{equation}
  \langle \hat{S}_\mathbf{n}^r \rangle_N^{|N,0 \rangle} = \sum_{k=0}^N
  (N - 2 k)^r \binom{N}{k}
  \sin^{2 k} \left ( \frac{\Theta}{2}  \right ) \cos^{2 (N - k)} \left (
  \frac{\Theta}{2} \right )  .
\end{equation}
The lowest-order polarization tensors are
\begin{eqnarray}
  \mathbf{T}^{(1,N)} (|N,0\rangle) =
  \left (
    \begin{array}{c}
      0 \\
      0 \\
      N
    \end{array}
  \right) ,
  \quad \quad
  \mathbf{T}^{(2,N)} (|N,0\rangle) =
  \left ( \begin{array}{ccc}
      N & i N & 0 \\
      -i N & N & 0 \\
      0 & 0 & N^2
    \end{array}
  \right ) , \\
  \mathbf{T}^{(3,N)} (|N,0\rangle) =
  \left ( \begingroup \everymath{\scriptstyle}
    \begin{array}{ccc}
      0 & 0 & N^2 \\
      0 & 0 & i N^2 \\
      N (N-2) & i N (N-2) & 0 \\  \hline
      0 & 0 & -i N^2 \\
      0 & 0 & N^2 \\
      -i N (N-2) & N (N-2) & 0 \\ \hline
      N^2 & i N^2 & 0 \\
      -i N^2 & N^2 & 0 \\
      0 & 0 & N^3 \end{array} \endgroup \right ) .
\end{eqnarray}
For $N > 0$, the SU(2) coherent states thus satisfy
$\mathbb{P}_S = 1$.

\subsection{Two-mode coherent states}

Since any pair of two-mode coherent states with the same average total
energy are related by an SU(2) transformation, it suffices to study
states of the form
\begin{equation}
  |\alpha,0 \rangle = e^{-|\alpha|^2/2} \sum_{N=0}^\infty
  \frac{\alpha^N}{\sqrt{N!}} \, |N,0 \rangle .
\end{equation}
The block-diagonal projection is clearly independent of the phase of
$\alpha$, and is given by a Poissonian mixture of SU(2) coherent
states that all belong to different excitation manifolds.  Hence, the
manifold-specific expectation values coincide with those of the SU(2)
coherent states, and the manifold-averaged Stokes momentum profiles
(\ref{SumN}) become
\begin{equation}
 \langle \hat{S}_\mathbf{n}^r \rangle^{|\alpha,0 \rangle} =
 \sum_{N=0}^\infty \frac{\bar{N}^N e^{-\bar{N}}}{N!} \,
 \langle \hat{S}_\mathbf{n}^r \rangle_N^{|N,0 \rangle} ,
\end{equation}
where $\bar{N} = |\alpha|^2$.  Using the corresponding tensor relation
(\ref{SumN}) and results for the states $|N,0 \rangle$, we easily
obtain
\begin{eqnarray}
  \fl \langle \hat{S}_\mathbf{n} \rangle^{|\alpha,0 \rangle} =
   \bar{N} n_3 ,
  \qquad
   \langle \hat{S}_\mathbf{n}^2 \rangle^{|\alpha,0 \rangle} =
   \bar{N} (1 + \bar{N} n_3^2) ,
  \qquad
   \langle \hat{S}_\mathbf{n}^3 \rangle^{|\alpha,0 \rangle} =
   \bar{N} n_3 (1 + 3 \bar{N} + \bar{N}^2 n_3^2) , \label{Scoh} \\
  \fl \mathbf{T}^{(1)} (|\alpha,0\rangle) =
   \left ( \begin{array}{c}
   0 \cr 0 \cr \bar{N}
   \end{array}
   \right ) ,
  \; \;
 \mathbf{T}^{(2)} (|\alpha,0\rangle) =
  \left ( \begin{array}{ccc}
  \bar{N} & i \bar{N} & 0 \\
   -i \bar{N} & \bar{N} & 0 \\
   0 & 0 & \bar{N} (\bar{N} + 1)
  \end{array}
  \right ) , \\
  \fl \mathbf{T}^{(3)} (|\alpha,0\rangle) =
  \left ( \begingroup \everymath{\scriptstyle} \begin{array}{ccc}
      0 & 0 & \bar{N} (\bar{N} + 1) \\
     0 & 0 & i \bar{N} (\bar{N} + 1) \\
   \bar{N} (\bar{N} - 1) & i \bar{N} (\bar{N} - 1) & 0 \\ \hline
      0 & 0 & -i \bar{N} (\bar{N} + 1) \\
     0 & 0 & \bar{N} (\bar{N} + 1) \\
    -i \bar{N} (\bar{N} - 1) & \bar{N} (\bar{N} - 1) & 0 \\ \hline
      \bar{N} (\bar{N} + 1) & i \bar{N} (\bar{N} + 1) & 0 \\
   -i \bar{N} (\bar{N} + 1) & \bar{N} (\bar{N} + 1) & 0 \\
   0 & 0 & \bar{N} (\bar{N}^2 + 3 \bar{N} + 1)
  \end{array} \endgroup \right ) .
\end{eqnarray}
In accordance with classical optics, $\mathbb{P}_S = 1$ for any
two-mode coherent state with a finite average photon number $\bar{N}$.
We note that when $\bar{N} \gg 1$, the lowest Stokes moments satisfy
$\left. \langle \hat{S}_\mathbf{n}^r \rangle^{|\alpha,0 \rangle}
\right|_{n_3=1} \approx (\bar{N} n_3)^r$. For $n_3 = 1$, the Stokes
moments (\ref{Scoh}) are directly given by the Poissonian photon
distribution, and the approximation corresponds to the classical
deterministic limit. The $n_3$ dependence of the approximation
describes the transmission through a classical beam splitter or
linear polarizer. In particular, Malus' law is obtained for $r = 1$.

\subsection{$|m,m \rangle$ states}

When considering the two-mode Fock states $|m,m \rangle$, which allow
for Heisenberg-limited interferometry~\cite{Holland}, we implicitly
treat all states obtained from these by SU(2) transformations.  The
latter states can be expressed as
\begin{eqnarray}
 \lefteqn{\hat{U} (\varphi,\vartheta,\xi) | m , m \rangle =
 \frac{\sin^m \vartheta}{m! \, 2^m} \sum_{k=0}^{2 m}
 \frac{2^k \sqrt{(2 m - k)! \, k!}}{\tan^k \vartheta}} & & \nonumber \\
 & & \times \sum_{j=0}^{\lfloor \frac{k}{2} \rfloor} \binom{m}{j}
 \binom{m - j}{j + m - k} \left( - \frac{\tan^2 \vartheta}{4} \right)^j
 | 2 m - k , k \rangle .
\end{eqnarray}
where $\lfloor x \rfloor$ denotes the largest integer that is smaller
than or equal to $x$. We have also assumed that the binomial
coefficients are defined through the gamma function, so that negative
integers are allowed as arguments. The Stokes moment profiles of the
states $|m,m \rangle$ are given by
\begin{equation}
  \langle \hat{S}_\mathbf{n}^r \rangle_{2 m}^{|m,m\rangle} = \left\{
    \begin{array}{ll}
      \displaystyle 2^r \sum_{j=0}^{r/2} [(2 j - 1)!!]^2 F (r , 2 j)
      \binom{m + j}{2 j} \sin^{2 j} \Theta , & r \ \mathrm{even,} \\
       0 , & r \ \mathrm{odd,}
    \end{array}
  \right.
\end{equation}
where $F (n, k)$ denotes the central factorial numbers of the second
kind (\ref{CentralFactorialNumbersDef}).  As both arguments are even
in our case, we have \cite{Butzer}
\begin{equation}
  F (r, 2 j) = 2 \sum_{k=1}^j \frac{(-1)^{j+k} k^r}{(j + k)! (j - k)!} .
\end{equation}
In particular, we get
\begin{eqnarray}
  \langle \hat{S}_\mathbf{n}^2
  \rangle_N^{|\frac{N}{2},\frac{N}{2}\rangle}
  & = & \frac{N (N + 2) \sin^2 \Theta}{2} , \label{S2mm} \\
  \langle \hat{S}_\mathbf{n}^4
  \rangle_N^{|\frac{N}{2},\frac{N}{2}\rangle} & = &
  N (N + 2) \sin^2 \Theta \, \frac{16 + 3 (N - 2) (N + 4) \sin^2
  \Theta}{8} . \label{S4mm}
\end{eqnarray}
The photon-number symmetry makes all odd-order Stokes profiles
vanish. This is in stark contrast to the states $|N,0 \rangle$ and
$|\alpha,0 \rangle$, and makes $\mathbb{P}_S = 0$.
However, the so-called hidden polarization of the $|m,m \rangle$
states appears in the even-order Stokes profiles.
As $\sin^2 \Theta = 1 - n_3^2$, these profiles
have common features with rotated ones for $|2 m,0 \rangle$.
This is in agreement with the fact that more elaborate
measurements than those considered here are required
to achieve Heisenberg resolution when employing the
$|m,m \rangle$ states \cite{Holland2}.

\subsection{Two-mode squeezed vacuum}

Using the process of spontaneous parametric down-conversion, one can
straightforwardly generate two-mode squeezed vacuum states.  These
have thermal photon-pair distributions and take the form
\begin{equation}
  | \Psi_\mathrm{TMSV} \rangle = \sum_{m=0}^\infty e^{i \phi_m}
  \sqrt{\frac{2 \bar{N}^m}{(2 + \bar{N})^{m+1}}} \, |m,m \rangle ,
\end{equation}
where $\bar{N}$ denotes the average number of photons. From
(\ref{SumN}), (\ref{S2mm}) and (\ref{S4mm}), we thus get
\begin{eqnarray}
  \langle \hat{S}_\mathbf{n}^2 \rangle^\mathrm{TMSV} & = &
  \frac{\bar{N} (2 \bar{N} + 3) \sin^2 \Theta}{2} , \\
  \langle \hat{S}_\mathbf{n}^4 \rangle^\mathrm{TMSV} & = &
  \bar{N} \sin^2 \Theta \, \frac{32 \bar{N} + 48 + 9 \,
  (8 \bar{N}^3 + 20 \bar{N}^2 + 10 \bar{N} - 5) \sin^2 \Theta}{8} .
\end{eqnarray}

\begin{figure}[t]
  \includegraphics[width=0.90\columnwidth]{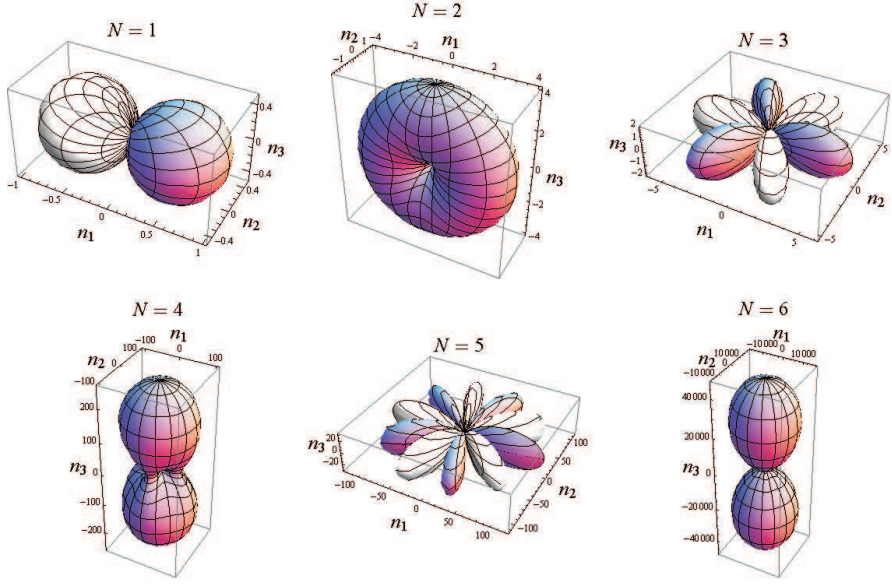}
  \caption{The Stokes moment profiles
  $\langle \hat{S}_\mathbf{n}^N \rangle_N$ for the $N$-photon
  NOON states with $1 \leq N \leq 6$. Dark and light surfaces
  indicate the directions for which
  $\langle \hat{S}_\mathbf{n}^N \rangle_N$ takes positive and
  negative values, respectively.}
  \label{Fig:NOONplots}
\end{figure}

\subsection{NOON states}

Finally, let us consider the NOON states $(| N , 0 \rangle + | 0 , N
\rangle)/\sqrt{2}$, where $N > 0$.  These are in some sense optimal
for interferometry~\cite{SoderholmPRA}. Their Stokes moment profiles
are found to be
\begin{equation}
  \fl \langle \hat{S}_\mathbf{n}^r \rangle_N^\mathrm{NOON} = \left\{
    \begin{array}{ll}
      0 , & r \ \mathrm{odd,} \ N \ \mathrm{even,} \\
      \displaystyle
      \frac{\cos (N \Phi) \sin^N \Theta}{4^{(N - 1)/2}}
      \sum_{k=0}^{\frac{N - 1}{2}} (N - 2 k)^r \binom{N}{k} (-1)^k , &
      r \ \mathrm{and} \ N \ \mathrm{odd,} \\
      \displaystyle \sum_{k=0}^N (N - 2 k)^r \binom{N}{k} \left[
        \cos^{2 k} \left ( \frac{\Theta}{2} \right ) \sin^{2 (N - k)}
        \left ( \frac{\Theta}{2} \right ) \right . & \\
      \displaystyle \left . + (-1)^k \cos (N \Phi) \cos^N  \left (
          \frac{\Theta}{2} \right ) \sin^N  \left ( \frac{\Theta}{2}
        \right ) \right ] , & r \ \mathrm{even.}
    \end{array} \right .
\end{equation}
We note that when $r$ is even and $N$ is odd, the effect of the second
term within square brackets vanishes.  For each of the first six NOON
states, we have plotted the Stokes moment profile $\langle
\hat{S}_\mathbf{n}^N \rangle_N$ in figure~\ref{Fig:NOONplots}.  In the
horizontal plane $n_3 = 0$, we have
\begin{equation}
  \left. \langle \hat{S}_\mathbf{n}^N \rangle_N^\mathrm{NOON}
  \right|_{\Theta=\pi/2} =  \left\{
    \begin{array}{ll}
      \displaystyle N! \cos (N \Phi) + 2^{N/2} Q_{N/2} ( N/2 ) , & N \
      \mathrm{even,} \\
      & \\
      \displaystyle N! \cos ( N \Phi) , & N \ \mathrm{odd,}
    \end{array}
  \right.
\end{equation}
where the polynomial \cite{Tuenter}
\begin{equation}
  Q_j (n) = 2^{j - 2 n} \sum_{k=0}^{2 n} \binom{2 n}{k} (n - k)^{2 j}
\end{equation}
satisfies $Q_0 (n) = 1$ and the recurrence relation $Q_{j+1} (n) = 2
n^2 Q_j (n) - n (2 n - 1) Q_j (n-1)$.  Hence, the de Broglie
wavelength of the NOON states, which scales as $N^{-1}$ and is the
reason for their superiority~\cite{Jacobson}, can be seen in these
measurements.

Let us now return to the pure two-photon states that satisfy
$\mathbb{P}_S = 0$.  These are given in (\ref{2phUnpol}) and are found
to be related to the two-photon NOON state by appropriate SU(2)
transformations according to
\begin{equation}
 | \Psi (a, \theta) \rangle = \hat{U} \left( \frac{\pi}{2} + \theta ,
 \arccos (\sqrt{2} a) , - \frac{\pi}{2} \right)
 \frac{| 2, 0 \rangle + | 0, 2 \rangle}{\sqrt{2}} .
\end{equation}
Hence, any Stokes moment profile of the state $| \Psi (a, \theta)
\rangle$ is related to the corresponding NOON profile by simple
rotations.  Since $\mathbf{R}_3 (\pi/2+\theta) \cdot \mathbf{R}_2
(\chi) \cdot \mathbf{R}_3 (-\pi/2) = \mathbf{R}_3 (\theta) \cdot
\mathbf{R}_1 (-\chi)$, $\langle \hat{S}_\mathbf{n}^2 \rangle_2^{| \Psi
  (a, \theta) \rangle}$ is obtained from $\langle \hat{S}_\mathbf{n}^2
\rangle_2^\mathrm{NOON}$ in figure~\ref{Fig:NOONplots} by applying a
rotation of $-\arccos (\sqrt{2} a)$ around $\mathbf{e}_1$ followed by
a rotation of $\theta$ around $\mathbf{e}_3$.  Since $\langle
\hat{S}_\mathbf{n} \rangle_2^{| \Psi (a, \theta) \rangle} = 0$ is
independent of $\mathbf{n}$, the states $| \Psi (a, \theta) \rangle$
lack first-order polarization structure.  However, they do all have a
second-order polarization structure.

\section{Conclusions}
\label{Sec:Conclusions}

Using expectation values of Stokes-operator products, we have
developed a systematic scheme for characterizing higher-order
polarization properties of two-mode quantized fields.
Polarization tensors and Stokes moment profiles were introduced
as two representations of the polarization information.
The latter show how passive interferometry affects the moments
of photon difference. This viewpoint was taken as
polarization properties of different states were compared.

Other possible representations of the polarization information
include central moments \cite{BjorkPRA}, quasi-probability
distributions \cite{Marquardt} and excitation-specific
generalized coherence matrices. Complete polarization
characterization requires the excitation manifolds to be
addressed separately. For situations where this cannot be
achieved, our characterization coincide with the one provided
by Klyshko's generalized coherence matrices \cite{Klyshko}.

Assuming ideal photon-number resolving detectors, we have
shown that it is possible to efficiently collect the data
through Stokes moment measurements in different directions.
In an experiment, it may be more practical to use more
measurement directions than the minimum required, but our
method should serve as a guide. In particular, we expect
the introduced moment components to be useful.
Another advantage of the described method is that it treats
the Stokes moments order by order. Hence, if only the first
few polarization orders are of interest,
it makes the measurements easier.

Since the different excitation manifolds are treated separately,
losses have drastic consequences in that
higher excitation manifolds then contribute to the lower ones.
Furthermore, whereas linear losses often model imperfections of
single-photon detectors well, photon-number resolving
detectors, which are required for full polarization
characterization, are more complex and may call for
nonlinear modeling.

On the other hand, the separation of data into excitation
manifolds and moment orders may be useful when developing
methods for efficient determination of polarization
characteristics. For example, one can take into account that
all state projections $\hat{\varrho}_N$ in the different
excitation manifolds must be physical states.
In this way, it should be possible to develop efficient
maximum likelihood methods similar to those regularly
employed in common quantum tomography.

\ack

Financial support from the Swedish Foundation for International
Cooperation in Research and Higher Education (STINT), the Swedish
Research Council (VR) through its Linn{\ae}us Center of Excellence
ADOPT and contract No.\ 621-2011-4575, the CONACyT (Grant No.
106525), the Spanish DGI (Grants FIS2008-04356 and FIS2011-26786), and
the UCM-BSCH program (Grant GR- 920992) is gratefully acknowledged.

\appendix
\section*{Appendix}
\setcounter{section}{1}

This appendix gives a derivation of the recurrence relation
(\ref{Srecurrence}), which involves central factorial numbers
\cite{Butzer}. We let the sets of non-negative and positive
integers be denoted as $\mathbb{N}_0$ and $\mathbb{N}_+$,
respectively. For $x \in \mathbb{R}$, the central factorial
of degree $n$ is defined by
\begin{equation}
 x^{[n]} = \left\{ \begin{array}{ll} 1 , & n = 0 , \\[.05in]
 x \displaystyle \prod_{k=2-n}^{n-2} \left( x + \frac{k}{2} \right)
 , \quad \quad \quad & n \in \mathbb{N}_+ . \end{array} \right.
 \label{CentralFactorialDef}
\end{equation}
The central factorial numbers of the first and second kind, $f (n,k)$
and $F (n,k)$ with $n,k \in \mathbb{N}_0$, respectively, are then
defined through the expansions
\begin{equation}
 x^{[n]} = \sum_{k=0}^n f (n,k) \, x^k , \quad \quad \quad x^n =
 \sum_{k=0}^n F (n,k) \, x^{[k]} . \label{CentralFactorialNumbersDef}
\end{equation}
We note that
\begin{eqnarray}
 f (n,0) = F (n,0) = \delta_{n,0} , \label{fn0} \\
 f (n,n) = F (n,n) = 1 , \quad \quad \quad n \in \mathbb{N}_+ ,
 \label{fnn}
\end{eqnarray}
where $\delta_{n,k}$ denotes the Kronecker delta.
For $x \in \mathbb{R}$, we clearly have
\begin{eqnarray}
 x^{[2 \nu]} = \prod_{j=0}^{\nu-1} (x^2 - j^2) , \quad \quad \quad
 \nu \in \mathbb{N}_+, \label{CFeven} \\
 x^{[2 \nu + 1]} = x \prod_{j=1}^\nu \left[ x^2 - \left( j -
 \frac{1}{2} \right)^2 \right] , \quad \quad \quad \nu \in
 \mathbb{N}_0 . \label{CFodd}
\end{eqnarray}
Consequently, both $f (n,k)$ and $F (n,k)$ vanish if one argument is
even and the other is odd.
For $m \in \mathbb{N}_0$, $\nu \in \mathbb{N}_+$, and $m < \nu$,
(\ref{CentralFactorialNumbersDef}), (\ref{fn0}) and (\ref{CFeven}) give
\begin{equation}
 m^{[2 \nu]} = \sum_{j=1}^\nu f (2 \nu, 2 j) \, m^{2 j} = 0 .
 \label{mIntegerEvenPowers}
\end{equation}
Factoring out an $m$ and making use of (\ref{fnn}), we obtain
\begin{equation}
 m^{2 \nu - 1} = - \sum_{j=1}^{\nu-1} f (2 \nu, 2 j) \, m^{2 j - 1} .
 \label{mIntegerOddPowers}
\end{equation}
Similarly, for $m , \nu \in \mathbb{N}_+$ and $m \leq \nu$, it follows
from (\ref{CFodd}) that
\begin{equation}
 (m - 1/2)^{[2 \nu + 1]} = \sum_{j=0}^\nu f (2 \nu + 1, 2 j + 1)
 (m - 1/2)^{2 j + 1} = 0 \label{mHalfintegerOddPowers}
\end{equation}
and
\begin{equation}
 \sum_{j=0}^\nu f (2 \nu + 1, 2 j + 1) (m - 1/2)^{2 j} = 0 .
\end{equation}
Hence, for any integer $m$ satisfying $1 - \nu \leq m \leq \nu$,
we have
\begin{equation}
 (m - 1/2)^{2 \nu} = - \sum_{j=0}^{\nu-1} f (2 \nu + 1, 2 j + 1)
 (m - 1/2)^{2 j} . \label{mHalfintegerEvenPowers}
\end{equation}

Now, consider an observable $\hat{A}$ in $d$-dimensional Hilbert
space with eigenvalues $\{ \lambda_j \}_{j=1}^d$.
In the eigenbasis, we then have
$\hat{A}^r = \mathrm{Diag} (\lambda_1^r , \lambda_2^r , \ldots , \lambda_d^r)$.
If all eigenvalues are integers and satisfy $|\lambda_k| < \nu$,
where $\nu \in \mathbb{N}_+$, (\ref{mIntegerOddPowers}) gives the
recurrence relation
\begin{equation}
 \hat{A}^{2 \nu - 1 + \mu} = - \sum_{j=1}^{\nu-1} f (2 \nu, 2 j) \,
 \hat{A}^{2 j - 1 + \mu} , \label{Aintegers}
\end{equation}
which is valid for any  $\mu \in \mathbb{N}_0$.
If all eigenvalues are half-integers and satisfy $|\lambda_k| \leq \nu - 1/2$,
where $\nu \in \mathbb{N}_+$,
it follows from Eq~(\ref{mHalfintegerEvenPowers}) that
\begin{equation}
 \hat{A}^{2 \nu + \mu} = - \sum_{j=0}^{\nu-1} f (2 \nu + 1, 2 j + 1) \,
 \hat{A}^{2 j + \mu} . \label{Ahalfintegers}
\end{equation}
Here, $\mu$ can take any integer value, since the inverse of $\hat{A}$ is
guaranteed to exist. We note that we can apply arbitrary unitary
transformations to both sides of (\ref{Aintegers}) and (\ref{Ahalfintegers}),
so they are valid in any basis.

Now, consider the angular momentum component $\hat{J}_3$ of a spin-$s$ particle.
The dimension of the corresponding Hilbert space is $d = 2 s + 1$ and
the eigenvalues are $- s ,- s + 1 , \ldots , s$.
For $s \in \mathbb{N}_+$, the magnitudes of all eigenvalues are integers
less than $s + 1$ and (\ref{Aintegers}) becomes
\begin{equation}
 \hat{J}_3^{2 s + 1 + \mu} = - \sum_{j=1}^s f (2 s + 2, 2 j) \,
 \hat{J}_3^{2 j - 1 + \mu} . \label{Jintegers}
\end{equation}
If $s$ is not an integer, all eigenvalues are half-integers with magnitudes
less than or equal to $s$ and (\ref{Ahalfintegers}) becomes
\begin{equation}
 \hat{J}_3^{2 s + 1 + \mu} = - \sum_{j=0}^{s-1/2} f (2 s + 2, 2 j + 1) \,
 \hat{J}_3^{2 j + \mu} . \label{Jhalfintegers}
\end{equation}
Taking the expectation value of the corresponding recurrence relations for
the Stokes operator $\hat{S}_3 = 2 \hat{J}_3$ in excitation manifold $N = 2 s$,
gives the desired result (\ref{Srecurrence}).


\newpage

\begin{thebibliography}{00}

\bibitem{Klyshko}
Klyshko D N 1997 \textit{Sov. Phys. JETP} \textbf{84} 1065

\bibitem{Muller}
M\"{u}ller Ch,  Stoklasa B, Klimov A B, Peuntinger Ch, Gabriel Ch,
\v{R}eh\'{a}\v{c}ek J, Hradil Z, Leuchs G, Marquardt Ch and
S\'anchez-Soto L L 2012 \NJP \textbf{14} 085002

\bibitem{Sergienko}
Jaeger G, Teodorescu-Frumosu M, Sergienko A,  Saleh B E A and  Teich M
C 2003 \PR A \textbf{67} 032307

\bibitem{BjorkPRA}
Bj\"{o}rk G, S\"{o}derholm J, Kim Y-S, Ra Y-S, Lim H-T, Kothe C, Kim Y-H, S\'anchez-Soto L L and Klimov A B 2012 \PR A \textbf{85} 053835

\bibitem{RaymerQCM} Raymer M G, Funk A C and McAlister D F 2000 \textit{Quantum Communication, Computing, and Measurement 2} ed Kumar P \etal (New York: Plenum) p~147

\bibitem{RaymerPRA} Raymer M G and Funk A 2000 \PR A \textbf{61} 015801

\bibitem{Karassiov} Karassiov V P 2005 \textit{J. Russ. Laser Res.} \textbf{26} 484

\bibitem{Carozzi} Carozzi T, Karlsson R and Bergman J 2000 \PR E \textbf{61} 2024

\bibitem{Setala} Set\"{a}l\"{a} T, Lindfors K, Kaivola M, Tervo J and Friberg A T 2004 \textit{Opt. Lett.} \textbf{29} 2587

\bibitem{Luis3D} Luis A 2005 \PR A \textbf{71} 063815

\bibitem{Stokes} Stokes G G 1852 \textit{Trans. Cambridge Philos. Soc.} \textbf{9} 399

\bibitem{Collett} Collett E 1970 \textit{Am. J. Phys.} \textbf{38} 563

\bibitem{Schwinger} Schwinger J 1965 \textit{Quantum Theory of Angular Momentum} ed Biedenharn L C and van Dam H (New York: Academic) p~229.

\bibitem{Yurke} Yurke B, McCall S L and Klauder J R 1986 \PR A \textbf{33} 4033

\bibitem{LuisPO} Luis A and S\'{a}nchez-Soto L L 2000 \textit{Prog. Opt.} \textbf{41} 421

\bibitem{BjorkOC} Bj\"{o}rk G, S\"{o}derholm J, S\'{a}nchez-Soto L L, Klimov A B, Ghiu I, Marian P and Marian T A 2010 \textit{Opt. Commun.} \textbf{283} 4440

\bibitem{Glauber} Glauber R J 1963 \PR \textbf{130} 2529

\bibitem{Korolkova} Luis A and Korolkova N 2006 \PR A \textbf{74} 043817

\bibitem{Newton} Newton R G and Young B 1968 \APNY \textbf{49} 393

\bibitem{Leonhardt} Leonhardt U 1996 \PR A \textbf{53} 2998

\bibitem{Weigert} Weigert S 2006 \textit{Int. J. Mod. Phys.} B \textbf{20} 1942

\bibitem{Schilling} Schilling U, von Zanthier J and Agarwal G S 2010 \PR A \textbf{81} 013826

\bibitem{Simon} Simon R and Mukunda N 1990 \PL A \textbf{143} 165

\bibitem{FejesToth} Fejes T\'oth L 1965 \textit{Acta Math. Acad. Sci. Hungar.} \textbf{16} 437

\bibitem{Conway} Conway J H, Hardin R H and Sloane N J A 1996 \textit{Exp. Math.} \textbf{5} 139

\bibitem{Bogdanov} Yu. I. Bogdanov, G. Brida, I. D. Bukeev, M. Genovese, K. S. Kravtsov, S. P. Kulik, E. V. Moreva, A. A. Soloviev, and A. P. Shurupov,
Phys. Rev. A \textbf{84}, 042108 (2011).

\bibitem{Butzer} Butzer P L, Schmidt M, Stark E L and Vogt L 1989 \textit{Numer. Funct. Anal. Optimiz.} \textbf{10} 419

\bibitem{Holland} Holland M J and Burnett K 1993 \PRL \textbf{71} 1355

\bibitem{Holland2} Holland M J and Burnett K 2004 \PRL \textbf{92} 209302

\bibitem{SoderholmPRA} S\"{o}derholm J, Bj\"{o}rk G, Tsegaye T and Trifonov A 1999 \PR A \textbf{59} 1788

\bibitem{Tuenter} Tuenter H J H 2002 \textit{Fibonacci Quarterly} \textbf{40} 175

\bibitem{Jacobson} Jacobson J, Bj\"{o}rk G, Chuang I and Yamamoto Y 1995 \PRL \textbf{74} 4835

\bibitem{Marquardt} Marquardt C, Heersink J, Dong R, Chekhova M V, Klimov A B, S\'{a}nchez-Soto L L, Andersen U L and Leuchs G 2007 \PRL \textbf{99} 220401

\end{thebibliography}
\end{document}